\documentclass[conference,10pt]{IEEEtran}

\usepackage{amsmath,amssymb,amsthm}
\usepackage{graphicx}
\usepackage{booktabs}
\usepackage{cite}
\usepackage{url}
\usepackage{fancyhdr}
\usepackage{xcolor}
\usepackage{algorithm}
\usepackage{algorithmic}
\usepackage{bm}
\usepackage{microtype}
\usepackage{enumitem}

\newcommand{\E}{\mathbb{E}}
\newcommand{\Pa}{\mathrm{Pa}}
\newcommand{\doOp}{\mathrm{do}}
\newcommand{\SCM}{\mathcal{M}}

\newcommand{\Vset}{\mathcal{V}}
\newcommand{\Uset}{\mathcal{U}}
\newcommand{\Fset}{\mathcal{F}}

\newcommand{\xbf}{\boldsymbol{x}}
\newcommand{\ubf}{\boldsymbol{u}}

\fancypagestyle{codefooter}{%
  \fancyhf{}%
  \fancyfoot[L]{%
    \ifnum\value{page}=4\relax
      \raisebox{0.1in}[0pt][0pt]{\footnotesize\textsuperscript{1}Code: \url{https://github.com/THE-TRAIN-LAB/H-SCM-scheduling}.}
    \fi
  }%
}

\begin{document}
\pagestyle{codefooter}

\title{From Confounding to Counterfactuals: A Cross-Layer Causal Model for RAN Scheduling}

\vspace{-6mm}\author{
  \IEEEauthorblockN{Artin Elhamirad\IEEEauthorrefmark{1},
  Christo Kurisummoottil Thomas\IEEEauthorrefmark{1}, Ali Tajer\IEEEauthorrefmark{2}}
  \IEEEauthorblockA{\small
\IEEEauthorrefmark{1}Department of Electrical \& Computer Engineering,
    Worcester Polytechnic Institute, Worcester, MA, USA\\
    \IEEEauthorrefmark{2}Department of Electrical, Computer, and Systems Engineering,\\
    Rensselaer Polytechnic Institute, Troy, NY, USA\\
    Email: \textit{aelhamirad@wpi.edu, cthomas2@wpi.edu, tajera@rpi.edu}
 \vspace{-5mm} }
\vspace{-5mm}}

\vspace{-4mm}\maketitle

\vspace{-3mm}\begin{abstract}
Data-driven control in radio access networks (RANs) selects resource
allocations using associations between control actions and key
performance indicators (KPIs) learned from network telemetry. Because
the shared radio environment influences both the scheduler's decisions
and the resulting performance, a bias known as confounding, such
associations overstate the effect of an allocation and misdirect
resources toward user equipment (UEs) that cannot benefit. Moreover, individual UEs respond heterogeneously to the
same decision, so even a corrected average effect does not identify
which UEs gain from additional resources. While causal inference has recently been applied to RANs, existing approaches do not model the cross-layer mechanisms
that connect an
allocation to its outcome. In this paper, a hierarchical structural
causal model (H-SCM) spanning the PHY, MAC, and network layers is
proposed that separates cell-load confounding from UE-specific
response heterogeneity. In
particular, a counterfactual scheduling problem is formulated in which
physical resource block (PRB) allocations are selected using per-UE
counterfactual goodput rather than observational conditional averages.
This problem is solved via a pipeline combining constrained causal
discovery, structural-equation fitting, and
abduction--action--prediction. Simulations demonstrate that naive
observational estimation overstates the full-PRB response by 73.3\%,
whereas adjustment for cell-load confounding reduces the relative error
to 9.8\%. Held-out interventions are predicted with a coefficient of
determination of 0.988. Counterfactual scheduling achieves
a mean relative loss in total cell goodput below 0.5\%
against the ground-truth optimum, compared with
2.0--2.9\% for a neural-network baseline.
\end{abstract}

\vspace{-2mm}
\section{Introduction}
\label{sec:intro}
\vspace{-2mm}

Network control in modern cellular radio access networks (RANs) relies on telemetry
and data-driven artificial intelligence (AI) models to support resource-allocation decisions such as
physical resource block (PRB) allocation among network slices and
scheduling-policy selection~\cite{polese2022coloran}. These models learn statistical associations between network states, control actions, and key performance indicators (KPIs), and use these to select the actions that maximize performance objectives. However, such \emph{statistical associations can be misleading} when the radio environment influences both the scheduler's decisions and the resulting KPIs.
For example, low cell load can simultaneously increase the PRBs available
to a user equipment (UE) and improve its link quality through reduced co-channel
interference. Consequently, the higher goodput observed at higher PRB
allocations may partly reflect lower cell load rather than
the effect of allocating more PRBs. This situation, in which a shared environmental variable influences both the control action and the outcome, is known as \emph{confounding} \cite{pearl2009causality}. A controller that ignores it can misattribute environmental effects to the action being optimized.

While this confounding can be corrected by accounting for the congestion state when estimating the allocation effect, such a correction does not by itself enable UE-specific control. The corrected estimate characterizes only the population-average response, whereas UEs differ in propagation, device, and traffic conditions. The same increase in PRB allocation may therefore yield a negligible gain for a channel-limited UE and a substantial gain for a PRB-limited UE. Per-UE resource allocation instead requires predicting the outcome that a specific UE would attain under an alternative allocation, given the conditions it actually experienced. Such a unit-specific prediction is known as a \emph{counterfactual} \cite{pearl2009causality}. This work therefore addresses the following question: \emph{How can a RAN estimate the causal effects of resource-allocation decisions from observational telemetry, free of confounding by the network state, and predict the counterfactual response of an individual UE to an alternative allocation?}

\vspace{-1mm}\subsection{Related Work}\vspace{-1mm}

Traditional resource-allocation approaches optimize weighted-throughput and proportional-fair objectives~\cite{bjornson2013optimal}. More recently, there has been growing interest in data-driven control for scheduling, resource allocation, and slice management through the RAN Intelligent Controller (RIC), including machine-learning-based xApps and reinforcement-learning methods~\cite{polese2022coloran} and \cite{lotfi2022oran}. Both classes of formulations optimize network-performance objectives without explicitly distinguishing statistical association from the interventional effect of a scheduling decision.

A principled approach to overcoming these limitations is causal inference, in which \emph{structural causal models} (SCMs) formalize the distinction between \emph{observational, interventional, and counterfactual queries}~\cite{pearl2009causality}, and causal structure can be learned from data~\cite{bib:peters2017}. In this framework, confounding biases the estimated effect of an allocation, whereas effect modification means that this effect differs across UEs or network states, and a variable such as congestion can play both roles~\cite{vanderweele2009distinction}.
In wireless networks, causal reasoning has been advocated as a foundation for AI-native systems, with applications including resilient beamforming, digital-twin modeling, and intent management~\cite{thomas2024causal}. CausalTune adjusts for environmental and operational confounders in RAN configuration tuning~\cite{xue2026causaltune}, while BLINC uses a large language model (LLM) to supply domain-knowledge priors for Bayesian structure learning in context-specific configuration~\cite{prasad2026blinc}. At the counterfactual level, conformal methods have been used to estimate the KPI that would have resulted from deploying a different RAN application~\cite{hou2025counterfactual}, and the interaction between a user, an LLM agent, and the network has been modeled as an SCM to evaluate alternative formulations of a user intent~\cite{farzaneh2026intent}.
These methods reason at the level of configurations, applications, or intents~\cite{xue2026causaltune,prasad2026blinc,hou2025counterfactual,farzaneh2026intent}, \emph{without modeling the cross-layer mechanisms that connect a resource-allocation decision to its outcome}. Moreover, each addresses either confounding or counterfactual evaluation in isolation, whereas UE-specific allocation requires both.

\vspace{-2mm}\subsection{Contributions}\vspace{-1mm}

The main contribution of this paper is a cross-layer causal
framework for RAN resource allocation that distinguishes
cell-load confounding from UE-specific response heterogeneity.
We formulate a \emph{counterfactual scheduling problem}
that allocates PRBs according to each UE's predicted
goodput gain and develop a
\emph{hierarchical structural causal model (H-SCM)}
including the physical (PHY), medium access control (MAC), and network layers to predict
these individual responses.
Simulations demonstrate that accounting for cell load
reduces the relative error in estimated goodput under
full-PRB allocation from $73.3\%$ to $9.8\%$.
Allocating PRBs based on each UE's predicted response
achieves a mean relative loss in total cell goodput
below $0.5\%$ against the ground-truth optimum.
\section{System Model and Scheduling Motivation}
\label{sec:probform}\vspace{-1mm}
\label{sec:sysmodel}

Consider a single-cell downlink RAN serving a set
$\mathcal{U}_s$ of $K$ UEs in each scheduling interval. We model the system across the PHY, MAC, and network layers, whose variables and mechanisms, such as link adaptation, hybrid automatic repeat request (HARQ) retransmission, and queueing, jointly determine the application-layer goodput. Cross-layer dependencies are modeled on a per-UE basis, while cell-level quantities such as the congestion state and the available PRB budget are shared among the UEs scheduled in the same interval.
Let $L\in[0,1]$ represent a cell-level congestion state summarizing
coupled serving-cell resource pressure and co-channel activity in the
surrounding radio environment.
As $L$ increases, the radio resources available for serving-cell scheduling decrease, and the co-channel interference experienced by the UE grows.  

\vspace{-2mm}\subsection{Wireless RAN System Model}
\label{subsec:sys}
\vspace{-1mm}
For notational simplicity, we omit the UE index $u$  in this  section.
In the \emph{PHY layer (L1)}, let $\gamma_0$ be the nominal
received signal-to-noise ratio (SNR), $f_{\mathrm D}$ the Doppler spread, and
$\tau_{\mathrm{rms}}$ the delay spread. The UE-specific shadow-fading
offset is modeled as
$\xi_{\mathrm{sf}}\sim\mathcal{N}(0,\sigma_{\mathrm{sf}}^2)$
~\cite{bib:3gpp38901}. The effective signal-to-interference-plus-noise ratio (SINR) can  be modeled as
\vspace{-2mm}\begin{equation}
  \gamma_{\mathrm{eff}}
  =
  \gamma_0
  + \xi_{\mathrm{sf}}
  - \beta_L L
  - \Delta_{\mathrm D}(f_{\mathrm D})
  - \Delta_{\tau}(\tau_{\mathrm{rms}})
  + \varepsilon_{\gamma}.
  \label{eq:sinr_load}
\vspace{-1mm}\end{equation}
Here, we model the penalties as
$\Delta_{\mathrm{D}}(f_{\mathrm{D}})
=a_{\mathrm{D}}\log_2(1+f_{\mathrm{D}}/f_0)$ and
$\Delta_{\tau}(\tau_{\mathrm{rms}})
=a_{\tau}\log_2(1+\tau_{\mathrm{rms}}/\tau_0)$,
where $a_{\mathrm{D}},a_{\tau}>0$ are penalty coefficients
in dB and $f_0,\tau_0>0$ are reference scales. The parameter $\beta_L>0$ scales the interference contribution of the load state, consistent with cellular load-coupling models~\cite{siomina2012load}, and $\varepsilon_{\gamma}$ is a residual noise term accounting for unmodeled effects on $\gamma_{\mathrm{eff}}$. 
The
block error rate (BLER) $p$ is modeled as  a function of $\gamma_{\mathrm{eff}}$ and $j^\star$. We select the highest-rate MCS satisfying the BLER target from the
3GPP MCS table~\cite{bib:3gpp38214}, which specifies the corresponding
modulation order $b_{\mathrm{mod}}$ and target code rate $R$, which determine the nominal PHY spectral efficiency,
$\eta_{\mathrm{PHY}}=b_{\mathrm{mod}}R$, measured in bits/s/Hz.

In the \emph{MAC layer (L2)}, we model the number of PRBs allocated to the UE by a load-dependent scheduling rule,
\vspace{-1mm}\begin{equation}
  T = \min\!\left\{
    N_{\mathrm{PRB}},
    \max\!\left\{
      1,\,
      \left\lfloor N_{\mathrm{PRB}}(1-\alpha_L L)\right\rfloor
      + \varepsilon_T
    \right\}
  \right\},
  \label{eq:prb_load}
\vspace{-1mm}\end{equation}
where $N_{\mathrm{PRB}}$ is the total number of available PRBs, $\alpha_L\in(0,1)$ controls the scheduler's sensitivity to load, and $\varepsilon_T$ captures slot-level scheduling variability. Given the allocated PRBs, the MAC-layer service rate is further shaped by retransmission overhead and device-specific radio-frequency (RF) variation. 
We approximate the overhead of HARQ retransmission procedures~\cite{bib:3gpp38321} using a truncated expected-retransmission surrogate.  Let $K_{\mathrm H}$ be the maximum number of retransmissions considered in the model, and $N_{\mathrm{retx}}\in\{0,\ldots,K_{\mathrm H}\}$  the number of HARQ retransmissions. Since $N_{\mathrm{retx}}$ is a nonnegative integer-valued random variable, its expectation equals the sum of its tail probabilities. Under the simplifying assumption that every transmission attempt fails independently with the same probability $p$, at least $k$ retransmissions are required exactly when the first $k$ attempts fail, so $\Pr(N_{\mathrm{retx}}\geq k)=p^k$ and
\vspace{-2mm}\begin{equation} \mathbb{E}\!\left[N_{\mathrm{retx}}\right] = \sum_{k=1}^{K_{\mathrm H}} \Pr\!\left(N_{\mathrm{retx}}\geq k\right) = \sum_{k=1}^{K_{\mathrm H}} p^k.\label{eq:harq_retx} \vspace{-2mm}\end{equation}
This expression approximates the retransmission overhead without explicitly modeling HARQ combining.

Device-specific RF variation, arising from differences in receiver hardware implementation such as front-end noise figure and component tolerances across UE models, is represented by $\xi_{\mathrm{hw}}\sim\mathcal{N}(0,\sigma_{\mathrm{hw}}^2)$. The MAC service rate is
$\mu=C_0T\eta_{\mathrm{PHY}}\max\{1+\kappa_{\mathrm{hw}}
\xi_{\mathrm{hw}},0.05\}/(1+\kappa_{\mathrm{HARQ}}
\mathbb{E}[N_{\mathrm{retx}}])$, where $C_0$ converts
to Mbps and the $\kappa$ coefficients scale hardware
variation and HARQ overhead. Although hardware impairments physically degrade the received
SINR~\cite{bjornson2013optimal}, we approximate their
net effect through an assumed MAC-layer multiplier, which
keeps the device-specific pathway separate from the propagation
pathway through $\xi_{\mathrm{sf}}$.

In the \emph{network layer (L3)}, traffic burstiness is characterized by $\xi_{\mathrm{tb}} \triangleq C_a^2 = \sigma_A^2/\mu_A^2$, the squared coefficient of variation of the UE's inter-arrival times $A$, where $\mu_A$ and $\sigma_A^2$ are the mean and variance of $A$. This is a standard variability measure in queueing theory~\cite{whitt1983qna}. Poisson arrivals have $\xi_{\mathrm{tb}}=1$, while $\xi_{\mathrm{tb}}>1$ indicates greater inter-arrival variability.
Using the $GI/G/1$ heavy-traffic approximation for mean waiting time~\cite{whitt1983qna}, assuming no service-time variability conditional on the modeled MAC rate, and applying Little's law, we model the mean queue backlog as
\vspace{-2mm}\begin{equation} Q = \left[ \frac{\rho^2\xi_{\mathrm{tb}}}{2(1-\rho)} +\varepsilon_3 \right]_+, \label{eq:queue} \vspace{-1mm}\end{equation} where $\rho=\min\{\lambda/\mu,\rho_{\max}\}$, $\lambda$ is the rate of the queued traffic component that loads the UE's buffer and sets its backlog through $\rho$, $\rho_{\max}<1$ caps utilization, $\varepsilon_3$ captures residual queueing variation, and $[x]_+=\max\{x,0\}$.
Further, we model the packet-loss probability, which captures residual decoding errors and queue buildup, as
\vspace{-2mm}\begin{equation}
  P_{\mathrm{loss}}
  =
  1-
  \left(1-p^{\,1+K_{\mathrm{H}}}\right)
  \exp\!\left(
    -\frac{
      [Q-q_{\mathrm{th}}]_{+}
    }{q_{\mathrm{scale}}}
  \right).
  \label{eq:pkt_loss}
\vspace{-2mm}\end{equation}
In \eqref{eq:pkt_loss}, the first factor is the probability that decoding succeeds within $K_{\mathrm{H}} + 1$ attempts under the same independent-attempt assumption as \eqref{eq:harq_retx}, and the second factor models the probability of avoiding queue-induced drops as decreasing smoothly once the backlog $Q$ exceeds a threshold $q_{\mathrm{th}}$. The parameter $q_{\mathrm{scale}}$ controls the rate of this decrease. The two loss mechanisms are treated as independent.
Finally, the application-layer goodput is $Y=\mu(1-P_{\mathrm{loss}})$, the goodput attainable by a saturated (full-buffer) flow, i.e., the service capacity net of residual decoding and buffer-overflow losses.

\vspace{-2mm}\subsection{Conventional Scheduling}
\label{sec:conv}\vspace{-2mm}

Using the per-UE model, we now consider scheduling across the UEs $\mathcal{U}_s$, where $T^{(u)}\!\in\!\mathcal{T}$ and $Y^{(u)}$ denote the PRB allocation and resulting goodput of UE $u$, respectively. The network state available to the scheduler when selecting the PRB allocation is
$\xbf^{(u)} \!=\! [L,\, \gamma_0^{(u)},\, f_{\mathrm{D}}^{(u)},\,
\tau_{\mathrm{rms}}^{(u)}]^{\mathsf{T}}$.
A conventional scheduler assigns each UE a scheduling
weight $w^{(u)}$, which encodes its priority under the operator's
fairness policy, and allocates PRBs from a total budget $B$
accordingly. A data-driven
weighted sum-rate scheduler can then be formulated as \cite{bjornson2013optimal}
\vspace{-2mm}\begin{equation}
\begin{aligned}
  \max_{\{t^{(u)}\}} \quad
  & \sum_{u\in\mathcal{U}_s} w^{(u)}
    \E\!\left[
      Y^{(u)}
      \mid T^{(u)}=t^{(u)},\xbf^{(u)}
    \right] \\
  \mathrm{s.t.}\quad
  & \sum_u t^{(u)}\leq B,
    \qquad t^{(u)}\in\mathcal{T}.
\end{aligned}
\label{eq:conv_opt}
\vspace{-2mm}\end{equation}

In \eqref{eq:conv_opt}, the conditional expectation is
estimated from historical telemetry. Since $\xbf^{(u)}$
includes cell load $L$, conditioning on this state blocks
the modeled confounding path. However, the tight coupling
between load and allocation limits observational support
for some candidate allocations. The conditional mean also
averages over unobserved UE-specific conditions. We propose
an H-SCM to infer these conditions from factual telemetry
and predict each UE's response to alternative allocations.


\vspace{-1mm}\section{Proposed H-SCM Framework for Causal Resource Allocation}
\label{sec:formulation}


\label{sec:hscm}

We model the cross-layer cause-effect relations  as an \emph{H-SCM}, defined as 
$\SCM=(\Uset,\Vset,\Fset,P_{\Uset})$~\cite{pearl2009causality}, where
each endogenous variable $V_i\in \mathcal{V}$ is determined by the structural equation
$V_i=f_i\!\left(\Pa(V_i),U_i\right),$
where $\Pa(V_i)$ represents its causal parents and $U_i\in \mathcal{U}$ is an exogenous
noise term.
The model is hierarchical because its endogenous variables, comprising
the PHY, MAC, and network quantities from
the effective SINR through the allocation and service rate to the
goodput, are partitioned by the three protocol layers,
$\Vset=\Vset^{(1)}\cup\Vset^{(2)}\cup\Vset^{(3)}$,
with no edges directed from higher to lower layers.
The  inputs to these mechanisms divide into context variables,
which remain fixed within a scheduling epoch, and mechanism-level noise,
which is redrawn in each realization. Let $\boldsymbol{z}$ collect the
context variables, namely the congestion state $L$, the radio-link
variables $\{\gamma_0, f_{\mathrm{D}}, \tau_{\mathrm{rms}}\}$, and the
unit-specific factors
$\{\xi_{\mathrm{sf}},\xi_{\mathrm{hw}},\xi_{\mathrm{tb}}\}$. The set
$\Uset$ contains the noise terms
$\{\varepsilon_\gamma, \varepsilon_T, \varepsilon_3\}$, and the remaining
mechanisms are deterministic given their parents.

\begin{figure}[t]
\centering
\includegraphics[width=0.9\columnwidth]{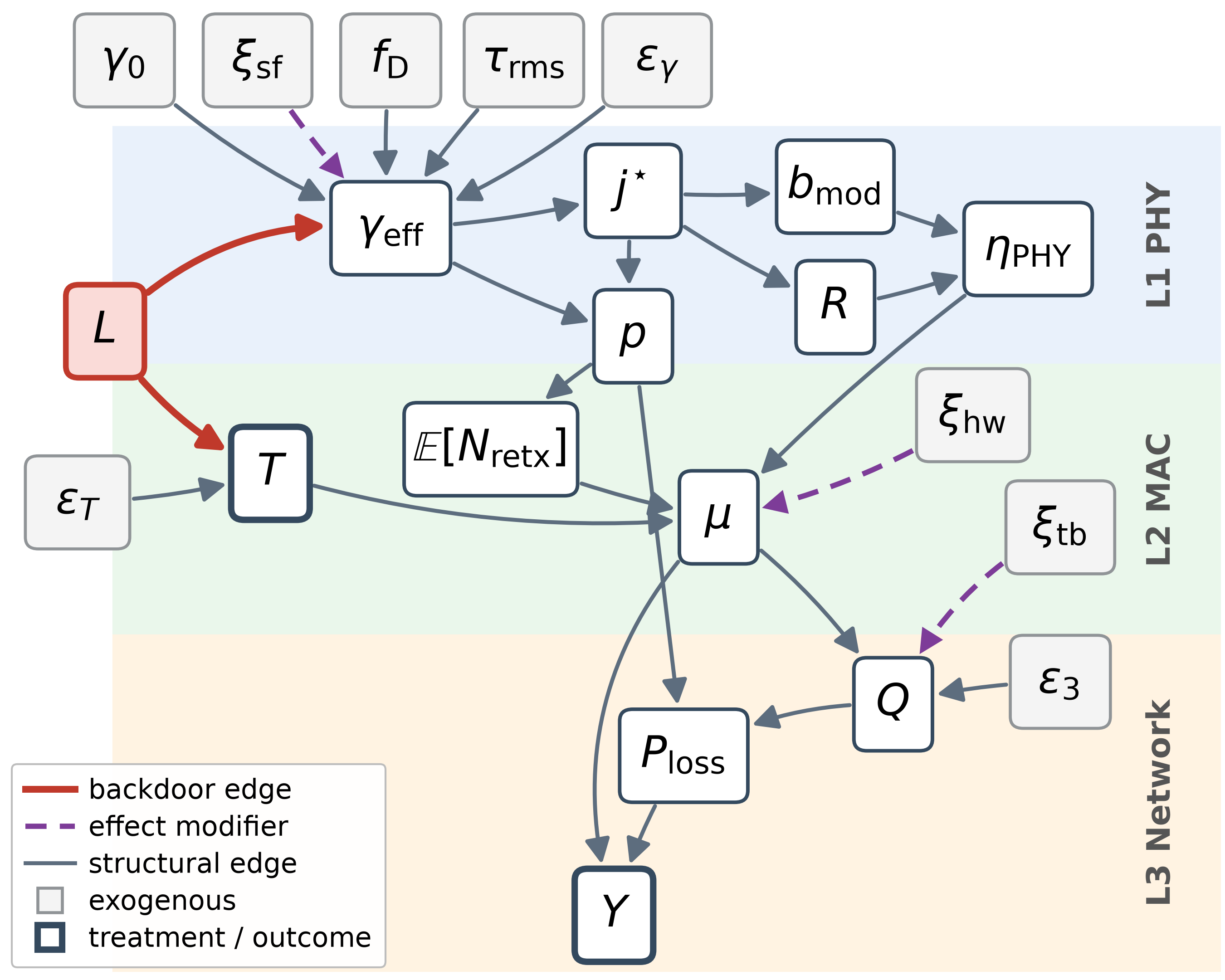}
\vspace{-3mm}\caption{Causal DAG spanning the PHY, MAC, and network layers. The DAG represents the simulated setting; operational RAN datasets may require additional nodes and dependencies.
}
\label{fig:dag}
\vspace{-6mm}\end{figure}

The structural equations induce the directed acyclic graph (DAG) in Fig.~\ref{fig:dag}.
Acyclicity follows from the layer partition together
with the feedforward processing order within each layer. The resulting observational distribution over context
$\boldsymbol z$ and endogenous-variable values
$\boldsymbol v=(v_i)_{V_i\in\Vset}$, including PRB
allocation and the resulting cross-layer KPIs, factorizes as
$
  P_{\SCM}(\boldsymbol z,\boldsymbol v)
  =
  P(\boldsymbol z)\prod_i P\!\left(v_i\mid\Pa(V_i)\right),$ where each conditional $P(v_i\mid\Pa(V_i))$ is induced by the mechanism
$f_i$ with $U_i$ marginalized out. For deterministic mechanisms it is a
point mass.

In the H-SCM, cell load $L$ is a common cause of the PRB allocation $T$ and effective SINR $\gamma_{\mathrm{eff}}$, creating the \emph{backdoor path} $T\leftarrow L\rightarrow\gamma_{\mathrm{eff}}\rightarrow\cdots\rightarrow Y$. Accounting for $L$ blocks this path when estimating the population
effect of PRB allocation. Herein, we define the do-operator $\doOp(T=t)$ that severs $L\rightarrow T$
but leaves $L\rightarrow\gamma_{\mathrm{eff}}$ active, since forcing the
scheduler does not alter the UE's interference environment. Cell load $L$ can also modify the goodput gain from additional
PRBs. In our DAG, the UE-specific factors
$\xi_{\mathrm{sf}}$, $\xi_{\mathrm{hw}}$, and
$\xi_{\mathrm{tb}}$ modify this gain without confounding
the allocation effect and are preserved during unit-level
counterfactual inference.

\vspace{-2mm}\subsection{Three Rungs of Causal Inference}
\label{sec:estimands}\vspace{-1mm}

The H-SCM supports observational, interventional, and counterfactual
queries about PRB allocation and goodput~\cite{pearl2009causality}.
At
the \emph{observational rung}, the mean goodput among UEs observed with
allocation $T=t$ is
\vspace{-2mm}
\begin{equation}
  \mu_{\mathrm{obs}}(t)=\E[Y\mid T=t],
  \label{eq:rung1}
\vspace{-2mm}\end{equation}
which, unlike the conditional mean in (6), does not condition on $L$ and is therefore confounded under the assumed DAG. At the
\emph{interventional rung}, the backdoor structure established previously
identifies the population response to setting $T\!=\!t$ by adjustment for
$L$ as
\vspace{-2mm}
\begin{equation}
  \mu_{\mathrm{int}}(t)
  =
  \E[Y\mid\doOp(T=t)]
  =
  \E_L\!\left[\E[Y\mid T=t,L]\right].
  \label{eq:rung2}
\vspace{-1mm}\end{equation}
Computing $\mu_{\mathrm{int}}(t)$ from observational data via
\eqref{eq:rung2} additionally requires the overlap condition $P(T=t\mid L)>0$ over the relevant load distribution~\cite{pearl2009causality}. Because the scheduler couples $T$ tightly to $L$, extreme allocations are rarely observed under unfavorable loads, so the controlled interventions in our evaluation provide direct coverage of these allocations.

At the \emph{counterfactual rung}, the predicted goodput of UE $u$ under allocation $t$ is $\hat{Y}_t^{(u)}=Y_t(\hat{\ubf}^{(u)})$, obtained by fixing the allocation to $T=t$ through $\doOp(T=t)$ and evaluating the fitted mechanisms with the UE-specific state $\hat{\ubf}^{(u)}$, inferred from factual telemetry and held fixed.

\vspace{-1mm}
\subsection{Counterfactual Scheduling Problem}
\label{sec:cf_sched}\vspace{-1mm}

For each UE $u$, we select an allocation $t^{(u)}$ to
maximize the weighted sum of counterfactual goodput
under a shared PRB budget. The counterfactual scheduling
problem is then
\vspace{-2mm}\begin{equation}
\begin{aligned}
  \max_{\{t^{(u)}\}} \quad
  & \sum_{u\in\mathcal{U}_s}
    w^{(u)}\,\hat{Y}_{t^{(u)}}^{(u)} \\
  \mathrm{s.t.}\quad
  & \sum_u t^{(u)} \leq B,
    \qquad t^{(u)}\in\mathcal{T}.
\end{aligned}
\label{eq:cf_opt}
\vspace{-2mm}\end{equation}
A conditional causal response can already support
UE-specific scheduling through the observed state
$\xbf^{(u)}$, but averages over unobserved conditions.
The objective in \eqref{eq:cf_opt} additionally uses
each UE's residual state inferred from factual telemetry
to predict its response to alternative allocations,
assuming that state persists. 
\vspace{-2mm}
\subsection{Proposed Solution via a Causal Inference Pipeline}\vspace{-2mm}
We solve~\eqref{eq:cf_opt} in three stages to predict
each UE's goodput under alternative allocations while
preserving its inferred conditions. Accordingly, we recover the causal
graph, fit the structural mechanisms, and infer each UE's
residual state from factual telemetry. The fitted model
then predicts the counterfactual goodput used in the
scheduling objective.
\vspace{-0mm}\subsubsection{Causal Discovery}
\label{sec:discovery}
In practice, the causal graph is not given and must be recovered from
data, a task known as causal discovery~\cite{bib:peters2017}. The
DAG in Fig.~\ref{fig:dag} serves as the simulator ground truth against
which discovery is evaluated, and the recovered graph supplies the
parent sets used for structural fitting and the causal order used for
counterfactual prediction. We use the Peter--Clark (PC) algorithm
~\cite{bib:peters2017} to learn an initial causal skeleton from
observable telemetry through conditional-independence testing.
We restrict edge directions using protocol-layer and within-layer
processing orders~\cite{bib:3gpp38214,bib:3gpp38321} and orient the
remaining edges with the Direct Linear Non-Gaussian
Acyclic Model (DirectLiNGAM)
~\cite{shimizu2011directlingam}.
DirectLiNGAM exploits non-Gaussianity to estimate a causal
ordering under a linear acyclic model. Since our mechanisms
are nonlinear, we use this ordering as a heuristic,
followed by interventional refinement.
Regression on pooled observational and
interventional data
proposes additional edges, which are admitted in significance order
while preserving acyclicity.

Candidate edges are then refined using simulator interventions
that hold selected mediators fixed. For an edge
$V_i\to V_j$, the mediator set $\boldsymbol M_{ij}$ contains variables
between $V_i$ and $V_j$ in protocol order that respond to interventions
on $V_i$ and are either ancestors of $V_j$ in the candidate graph or
children of $V_i$ connected by previously retained edges. These variables are
held fixed jointly, yielding the contrast
\begin{equation}
\begin{aligned}
  \Delta_{ij}
  ={}&\E\!\left[V_j\mid\doOp(V_i=x_1),
       \doOp(\boldsymbol M_{ij}=\boldsymbol m)\right] \\
     &-\E\!\left[V_j\mid\doOp(V_i=x_0),
       \doOp(\boldsymbol M_{ij}=\boldsymbol m)\right].
\end{aligned}
\label{eq:mediator}
\end{equation}
The two arms share identical exogenous draws. A candidate edge is retained if an effect remains significant
after mediator blocking, but unblocked indirect paths
can prevent interpreting it as a direct edge. Mediator selection uses the candidate graph and
protocol order, without consulting the ground-truth edges.

\subsubsection{Structural Equation Fitting}

For each endogenous variable $V_i$, its structural mechanism is estimated
from its causal parents as
\vspace{-2mm}\begin{equation}\vspace{-2mm}
  \hat{f}_i = \arg\min_f \frac{1}{N}\sum_{n=1}^{N}
  \left(v_i^{(n)}-f\!\left(\Pa(v_i^{(n)})\right)\right)^2.
  \label{eq:fitting}
\vspace{-1mm}\end{equation}
We fit each structural mechanism with a
gradient boosting machine (GBM) using regression-tree base learners~\cite{friedman2001greedy}, which capture nonlinear dependencies and interactions among its tabular parent variables.
Telemetry fits use observable parents in the recovered
graph.
Cross-fitting~\cite{chernozhukov2018dml}
computes each UE's residual using a model trained without 
that UE.

\subsubsection{Counterfactual Inference}
\label{sec:cf_alg}
Counterfactual predictions follow three
steps~\cite{pearl2009causality}. First, \emph{abduction} estimates each UE's residual state
$\hat{\ubf}^{(u)}$ from factual telemetry using the assumed
mechanism-specific noise rules.
Its counterfactual interpretation assumes correctly specified
parent sets and noise forms, accurate mechanism fits, and
exogenous conditions that remain fixed under the intervention.
The inferred residuals approximate these conditions but
also contain model and estimation error.
Second, \emph{action} replaces the PRB-allocation mechanism
with the assignment $T=t$.
Finally, \emph{prediction} evaluates the fitted GBM mechanisms
in topological order, using updated parent values and
applying each UE's abduced residual according to the
mechanism-specific rule.        
\setlength{\textfloatsep}{4pt plus 1pt minus 1pt}


\vspace{-1mm}\section{Simulation Results and Analysis}
\label{sec:system}\vspace{-1mm}

We evaluate the proposed framework in a single-cell
downlink RAN simulation using Sionna~1.2 for the physical
layer~\cite{hoydis2022sionna}\textsuperscript{1}. We use the 3GPP clustered
delay line model C (CDL-C) channel at $2.6$~GHz, with a fixed
delay spread of $100$~ns and a UE speed of
$3$~m/s~\cite{bib:3gpp38901}. The simulated MCS-specific BLER curves determine the
highest-rate MCS satisfying the BLER target at each
UE's effective SINR.
The resulting MCS and BLER values feed the MAC and
network-layer equations in Section~\ref{sec:probform}
to compute service rate, queue backlog, packet loss,
and goodput.
Table~\ref{tab:sim_params} lists the key simulation parameters.

\begin{table}[t]
\centering
\caption{Key Simulation Parameters}\vspace{-3mm}
\label{tab:sim_params}
\footnotesize
\setlength{\tabcolsep}{3pt}
\begin{tabular}{@{}p{0.40\columnwidth}
p{\dimexpr0.60\columnwidth-2\tabcolsep\relax}@{}}
\toprule 
\textbf{Parameter} & \textbf{Value} \\
\midrule
OFDM grid
& $76$ subcarriers, $15$~kHz spacing, $14$ symbols \\

Coding and MCS set
& LDPC with $512$ information bits; $12$ MCSs \\

BLER target / HARQ limit
& $0.1$ / $3$ retransmissions \\

PRB and load parameters
& $N_{\mathrm{PRB}}=100$, $\alpha_L=0.58$,
  $\beta_L=18$~dB \\

Queue loading rate $\lambda$ / utilization cap
& $5.5$~Mbps / $\rho_{\max}=0.95$ \\

Queue-loss parameters
& $q_{\mathrm{th}}=5$, $q_{\mathrm{scale}}=10$ \\
\bottomrule
\end{tabular}
\end{table}

Using data generated by combined PHY, MAC, and
network-layer simulation, we evaluate causal discovery on the $16$ observable telemetry
variables, connected by $21$ of the complete DAG's $27$ edges.
The remaining six nodes are the latent factors $\{\xi_{\mathrm{sf}},\xi_{\mathrm{hw}},\xi_{\mathrm{tb}}\}$ and noise terms $\{\varepsilon_\gamma,\varepsilon_T,\varepsilon_3\}$, which are unavailable in the assumed telemetry. Their outgoing edges are therefore excluded from discovery and its evaluation, while their effects remain present in the generated observations.
Each discovery run uses $500$ observational and $2940$ interventional samples, with an additional $2000$
paired UE realizations per mediator-testing contrast described in
Section~\ref{sec:discovery}.
Across $50$ independent realizations, the proposed pipeline
recovers $19$ of the $21$ observable edges without false
positives, achieving precision $1.000$, recall $0.905$,
and F1 score $0.950$. The interventional regimes and the mediator-blocking contrasts set internal variables that only a simulator can control, which is a limitation of the discovery stage, whereas fitting for the counterfactual
(CF) queries and held-out
validation uses observational telemetry only.

After recovering the causal graph, we fit the SCM according to
\eqref{eq:fitting} using $500$ observational
samples.
Each GBM uses $200$ trees, a maximum depth of $4$,
and a learning rate of $0.05$.
Prediction accuracy is measured by the coefficient of determination
($R^2$), the fraction of ground-truth outcome variance explained by the
predictions. For counterfactual prediction, residuals are additive for most mechanisms, multiplicative for the MAC service rate, and rescaled only in the traffic-dependent component for queue backlog. For CF-2 (hardware normalization), the MAC residual multiplier is set to one, representing nominal hardware under the assumed decomposition. On $5000$ held-out UEs, the per-UE counterfactual
predictions achieve $R^2$ values of $0.979$, $0.988$, and $0.994$
against the simulator's per-UE outcomes under $\doOp(T=50)$,
$\doOp(T=100)$, and hardware normalization, respectively.

\begin{figure}[t]
\centering
\includegraphics[width=\columnwidth]{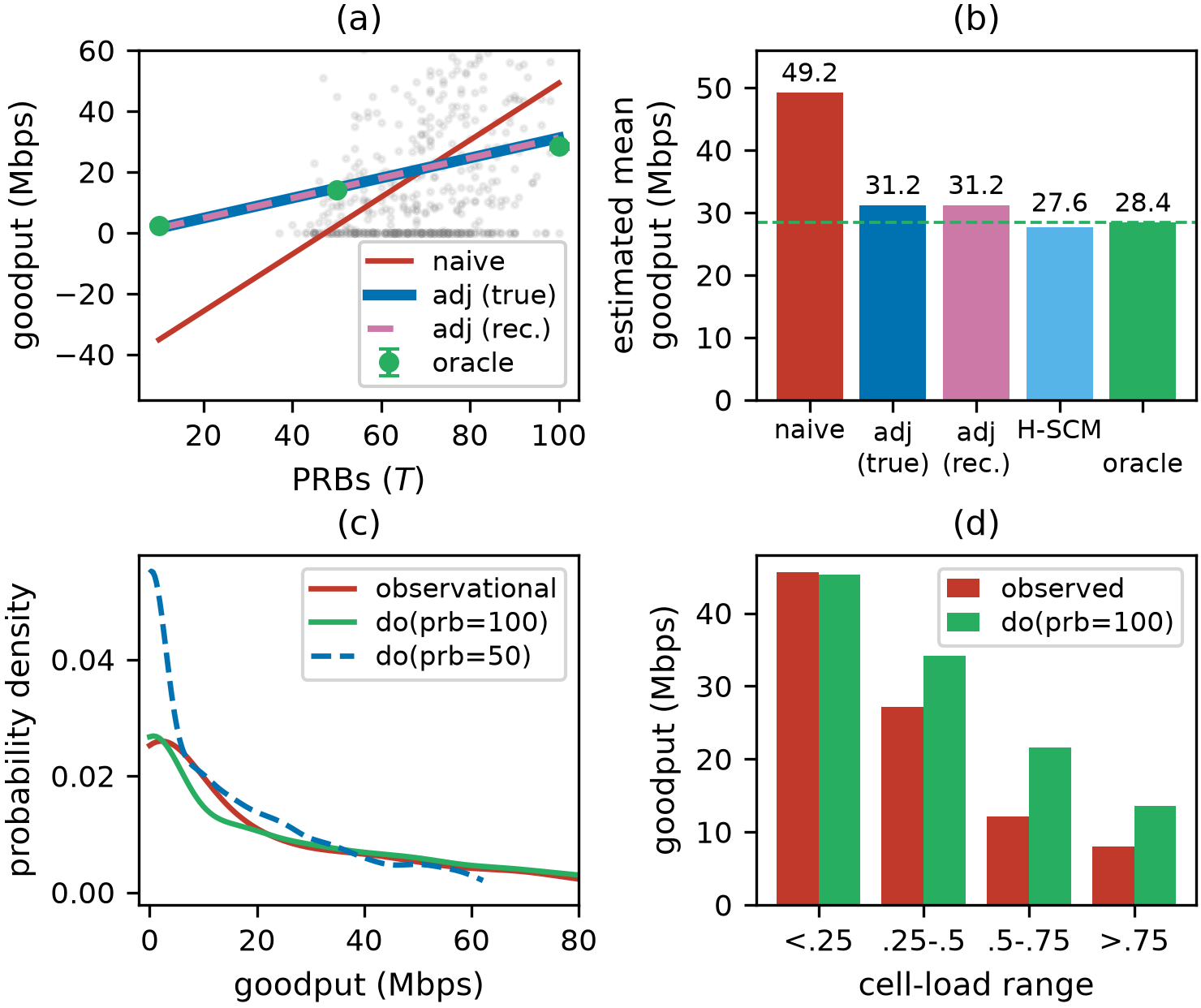}
\vspace{-8mm}\caption{Confounding analysis.
(a) Goodput versus PRB allocation.
(b) Full-PRB goodput estimates compared with
interventional ground truth.
(c) Goodput distributions.
(d) Mean goodput by cell-load range.}
\label{fig:confound}
\end{figure}

Fig.~\ref{fig:confound} compares goodput predictions from
naive ordinary least squares (OLS) regression of $Y$ on
PRB allocation $T$ alone, confounding-adjusted OLS
regression of $Y$ on both $T$ and cell load $L$, and
the H-SCM, against ground-truth goodput from simulator interventions.
In panels (a) and (b), naive OLS regression predicts
a mean goodput of $49.20$~Mbps at $T=100$, exceeding the
simulator's ground-truth interventional mean of
$28.39$~Mbps by $73.3\%$.
This naive estimate omits $L$, unlike the conventional
scheduler in~\eqref{eq:conv_opt}.
Including cell load $L$ in the regression and averaging
over the observed load distribution reduces the
predicted mean goodput to $31.18$~Mbps and the relative
error to $9.8\%$.
The recovered and reference graphs yield identical
adjusted estimates because the two edges missed during
causal discovery do not change the required adjustment
variable $L$.
In panel (b), averaging cross-fitted H-SCM counterfactual
predictions over the $500$ observational UEs gives
$27.64$~Mbps, a $2.7\%$ relative error against the
simulator mean over $5000$ held-out UEs.
Panels (c) and (d) show the goodput distributions and
responses across cell-load ranges.
The mean interventional goodput increases from
$2.42$ to $14.12$ and $28.39$~Mbps as the allocation
increases from $10$ to $50$ and $100$ PRBs.
However, fixing the allocation leaves the effect
of load on link quality active, so performance
still differs across load conditions.

\begin{figure}[t]
\centering
\includegraphics[width=\columnwidth]{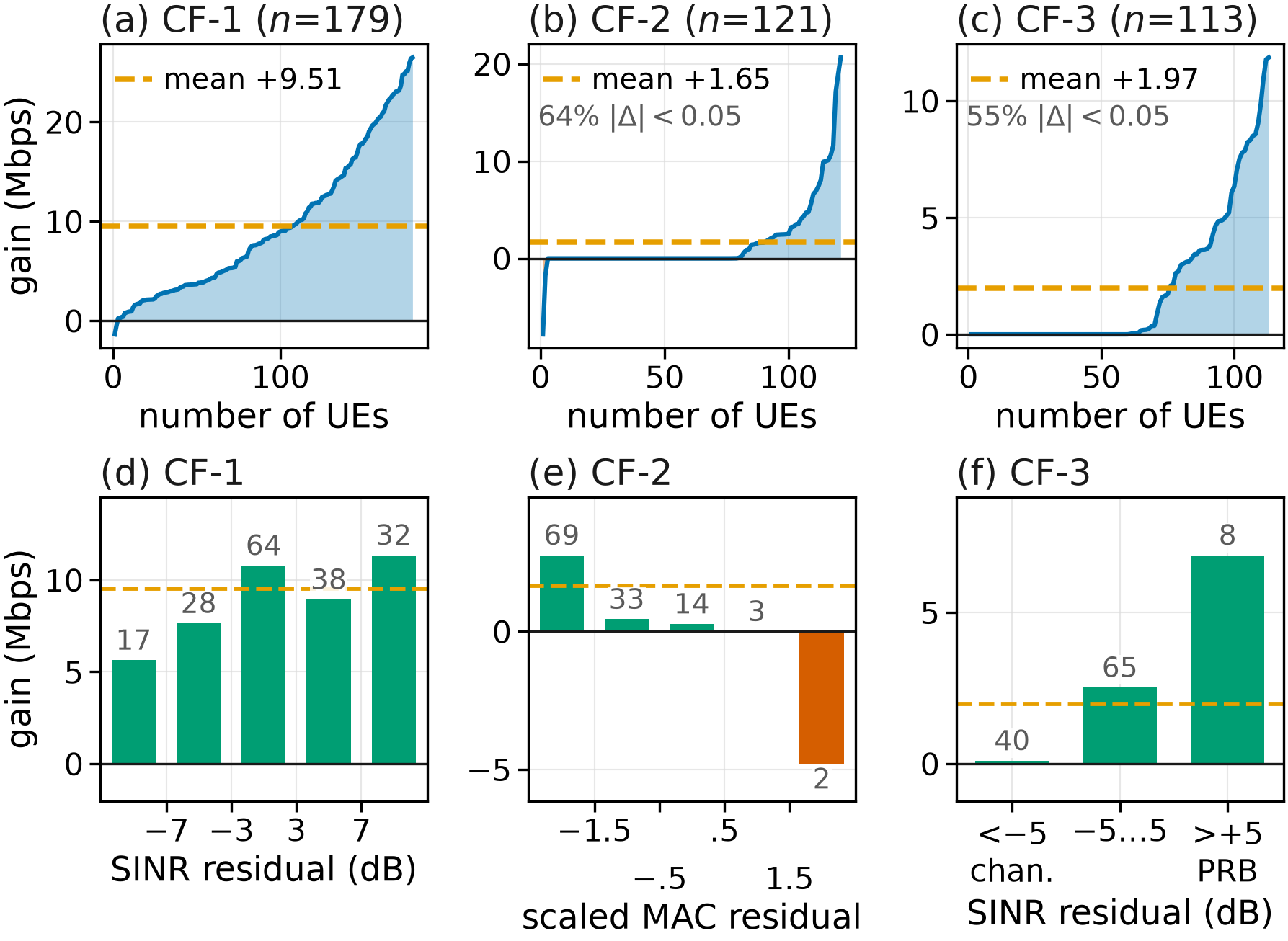}
\vspace{-8mm}\caption{Counterfactual goodput gains:
(a)--(c) sorted UE gains, where $n$ is the number of UEs selected
by each query, and (d)--(f) mean gains by inferred residual range.
Dashed lines show the mean gain. Bar labels give UE counts.}
\label{fig:cf_results}
\vspace{-1mm}
\end{figure}

In Fig.~\ref{fig:cf_results}, we show the results of three
CF queries. CF-1 and CF-3 evaluate the prediction
$\hat{Y}^{(u)}_{t}$ that constitutes the objective of \eqref{eq:cf_opt}
at the fixed allocation $t=100$, whereas CF-2 evaluates hardware normalization as a
model-based simulator diagnostic.
\emph{CF-1}, shown in panels (a) and (d), asks \emph{what goodput the
UEs with moderate factual goodput, $5\leq Y\leq40$~Mbps, would have
attained had they received the full PRB allocation $\doOp(T=100)$.}
This criterion selects $179$ UEs, whose mean goodput
increases from $19.53$ to $29.04$~Mbps,
an improvement of $48.7\%$.
Under full-PRB allocation, the mean increase in goodput relative to factual values is $5.65$~Mbps for UEs in the most channel-limited residual range and $11.35$~Mbps
for those in the most favorable range.
\emph{CF-2} asks \emph{what goodput the UEs underperforming
their peers would have attained had their MAC residual
multiplier been set to one.}
We select UEs whose MAC throughput is below $95\%$
of the median among peers with the same MCS and PRB
decile. Panel (b) shows the sorted per-UE goodput
changes. Panel (e) shows the mean predicted goodput change across
scaled MAC-residual ranges. Mean predicted goodput
increases from $14.15$ to $15.80$~Mbps ($11.7\%$),
with gains concentrated among UEs with a negative scaled MAC residual, i.e., a residual multiplier below one; those with a positive residual lose goodput.
\emph{CF-3}, shown in panels (c) and (f), asks \emph{what goodput the UEs
selected by the joint criteria of nominal SNR $\gamma_0\in[10,20]$~dB
and factual goodput $Y<15$~Mbps would have attained under
$\doOp(T=100)$.}
These criteria select $113$ UEs, whose mean goodput
increases from $3.55$ to $5.52$~Mbps,
an improvement of $55.6\%$.
However, UEs in the channel-limited residual range
gain only $0.11$~Mbps on average, compared with
$6.88$~Mbps for UEs in the favorable, PRB-limited range.
These differences show that the mean gain alone
cannot identify which UEs benefit from additional PRBs.

\begin{figure}[t]
 \vspace{-4mm}   \centering
    \includegraphics[width=\linewidth]{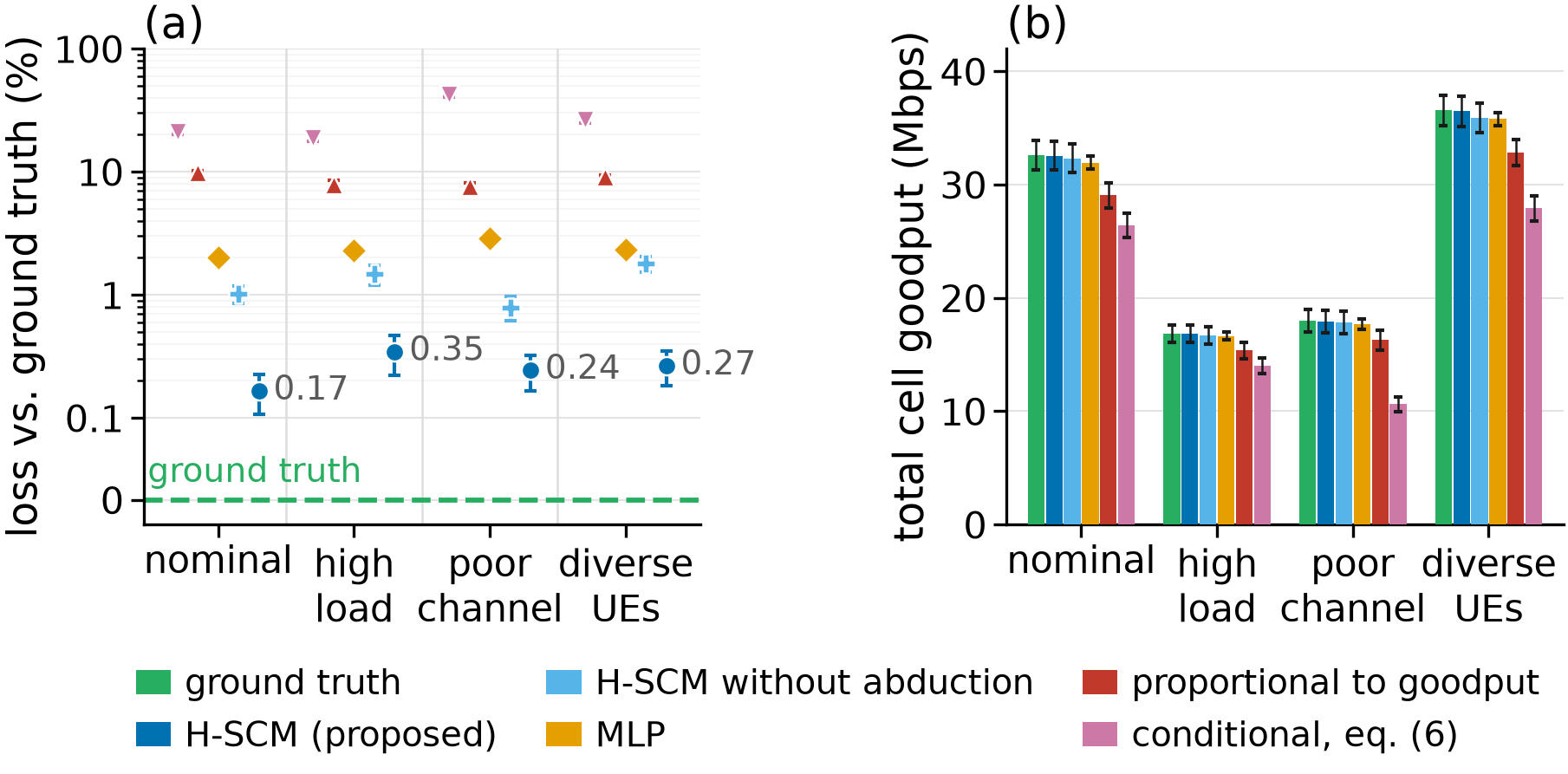}
   \vspace{-8mm} \caption{Scheduling performance across four environments:
    (a) loss relative to the ground-truth optimum and (b) realized cell goodput. Error bars show approximate $95\%$ confidence intervals
over $50$ realizations.}
    \label{fig:alloc}
\end{figure}

Fig.~\ref{fig:alloc} evaluates~\eqref{eq:cf_opt} for four UEs
($K=4$) sharing the load-dependent PRB budget derived
from~\eqref{eq:prb_load}, with at least $10$ PRBs per UE.
Training data comprise two scheduling epochs from $500$
four-UE cells in the nominal environment, first under equal
allocation and then a random feasible split.
\mbox{H-SCM} uses $4000$ unpaired rows. For scheduling, we fit the
MAC mechanism to $\mu/T$ using its remaining observable
parents and reconstruct its prediction by multiplying
by the candidate PRB allocation. An MLP with two
hidden layers of $64$ ReLU units each uses $2000$ pairs
to predict second-epoch goodput from full first-epoch
telemetry and candidate allocation.
Two reference policies are the same \mbox{H-SCM} without abduction,
setting residuals to their population defaults, and the
conventional scheduler of~\eqref{eq:conv_opt}, fitted as a GBM
of goodput on allocation and context $\xbf^{(u)}$ using the
same rows.
All policies are tested without refitting across four
environments over $50$ realizations. Relative to nominal conditions, high load raises mean $L$
from $0.50$ to $0.71$ and traffic-burstiness log-variance
from $0.12$ to $0.30$. Poor channel lowers the maximum
SNR from $28$ to $16$~dB, raises minimum Doppler/delay
spread from $5$~Hz/$30$~ns to $100$~Hz/$165$~ns,
and increases shadowing standard deviation from $7$
to $9$~dB. The diverse-UE environment increases hardware-factor
standard deviation from $1.5$ to $3$, shadowing
standard deviation from $7$ to $10$~dB, and
traffic-burstiness log-variance from $0.12$ to $0.40$.
Panel~(a) reports mean relative loss in realized total cell
goodput against the ground-truth optimum.
Loss is $0.17$--$0.35\%$ for \mbox{H-SCM},
$0.8$--$1.8\%$ without abduction, $2.0$--$2.9\%$ for MLP, $8$--$10\%$ for
goodput-proportional allocation, and $19$--$42\%$
for~\eqref{eq:conv_opt}.
Abduction reduces loss in every environment with identical
mechanisms.
Panel~(b) confirms that \mbox{H-SCM} closely approaches
ground-truth optimal goodput.

\vspace{-2mm}
\section{Conclusion}
\label{sec:conclusion}

In this paper, we have developed an H-SCM for cross-layer causal reasoning in wireless RANs. We have combined causal discovery, structural-equation fitting, and counterfactual inference to evaluate resource-allocation decisions. Simulation results have shown that accounting for cell-load confounding improves estimates of the effect of PRB allocation on goodput. Unit-level analysis has further shown that the same allocation can produce substantially different gains across UEs. 
Together, these results show that confounding adjustment and counterfactual inference address different needs in causality-aware RAN control. 
Future work will consider interventions that can be actuated on an operational RAN. We will also validate the recovered structure on Open RAN (O-RAN) traces.
\vspace{-2mm}\bibliographystyle{IEEEtran}
\def\baselinestretch{0.85}
\bibliography{refs}

@book{pearl2009causality,
  author    = {Pearl, Judea},
  title     = {Causality: Models, Reasoning, and Inference},
  edition   = {2},
  publisher = {Cambridge University Press},
  address   = {Cambridge, U.K.},
  year      = {2009},
  isbn      = {978-0-521-89560-6}
}

@book{bib:peters2017,
  author    = {Peters, Jonas and Janzing, Dominik and Sch{\"o}lkopf, Bernhard},
  title     = {Elements of Causal Inference: Foundations and Learning Algorithms},
  publisher = {MIT Press},
  address   = {Cambridge, MA, USA},
  year      = {2017},
  isbn      = {978-0-262-03731-0}
}

@article{shimizu2011directlingam,
  author  = {Shimizu, Shohei and Inazumi, Takanori and Sogawa, Yasuhiro and
             Hyv{\"a}rinen, Aapo and Kawahara, Yoshinobu and Washio, Takashi and
             Hoyer, Patrik O. and Bollen, Kenneth},
  title   = {{DirectLiNGAM}: A Direct Method for Learning a Linear
             Non-{G}aussian Structural Equation Model},
  journal = {Journal of Machine Learning Research},
  volume  = {12},
  pages   = {1225--1248},
  year    = {2011}
}

@article{friedman2001greedy,
  author  = {Friedman, Jerome H.},
  title   = {Greedy Function Approximation: A Gradient Boosting Machine},
  journal = {The Annals of Statistics},
  volume  = {29},
  number  = {5},
  pages   = {1189--1232},
  year    = {2001},
  doi     = {10.1214/aos/1013203451}
}

@article{chernozhukov2018dml,
  author  = {Chernozhukov, Victor and Chetverikov, Denis and Demirer, Mert and
             Duflo, Esther and Hansen, Christian and Newey, Whitney and
             Robins, James},
  title   = {Double/Debiased Machine Learning for Treatment and Structural Parameters},
  journal = {The Econometrics Journal},
  volume  = {21},
  number  = {1},
  pages   = {C1--C68},
  year    = {2018},
  doi     = {10.1111/ectj.12097}
}

@article{thomas2024causal,
  author  = {Christo Kurisummoottil Thomas and Christina Chaccour and Walid Saad and M{\'e}rouane Debbah and Choong Seon Hong},
  title   = {Causal Reasoning: Charting a Revolutionary Course for Next-Generation {AI}-Native Wireless Networks},
  journal = {IEEE Vehicular Technology Magazine},
  volume  = {19},
  number  = {1},
  pages   = {16--31},
  year    = {2024},
  doi     = {10.1109/MVT.2024.3359357}
}

@article{bjornson2013optimal,
  author    = {Bj{\"o}rnson, Emil and Jorswieck, Eduard},
  title     = {Optimal Resource Allocation in Coordinated
               Multi-Cell Systems},
  journal   = {Foundations and Trends in Communications and
               Information Theory},
  volume    = {9},
  number    = {2--3},
  pages     = {113--381},
  year      = {2013},
  publisher = {Now Publishers},
  doi       = {10.1561/0100000069}
}

@inproceedings{lotfi2022oran,
  author    = {Lotfi, Fatemeh and Semiari, Omid and Afghah, Fatemeh},
  title     = {Evolutionary Deep Reinforcement Learning for Dynamic
               Slice Management in {O-RAN}},
  booktitle = {Proc. of the 2022 {IEEE} Globecom Workshops ({GC Wkshps})},
  address   = {Rio de Janeiro, Brazil},
  month     = dec,
  pages     = {227--232},
  year      = {2022},
  doi       = {10.1109/GCWkshps56602.2022.10008614}
}

@article{polese2022coloran,
  title     = {{ColO-RAN}: Developing Machine Learning-Based {xApps} for Open {RAN} Closed-Loop Control on Programmable Experimental Platforms},
  author    = {Polese, Michele and Bonati, Leonardo and D'Oro, Salvatore and Basagni, Stefano and Melodia, Tommaso},
  journal   = {{IEEE} Transactions on Mobile Computing},
  volume    = {22},
  number    = {10},
  pages     = {5787--5800},
  year      = {2023},
  doi       = {10.1109/TMC.2022.3188013}
}

@article{siomina2012load,
  author    = {Siomina, Iana and Yuan, Di},
  title     = {Analysis of Cell Load Coupling for {LTE} Network Planning and Optimization},
  journal   = {{IEEE} Transactions on Wireless Communications},
  volume    = {11},
  number    = {6},
  pages     = {2287--2297},
  year      = {2012},
  doi       = {10.1109/TWC.2012.051512.111532}
}

@article{hoydis2022sionna,
  author    = {Hoydis, Jakob and Cammerer, Sebastian and
               {Ait Aoudia}, Fay{\c{c}}al and Vem, Avinash and
               Binder, Nikolaus and Marcus, Guillermo and
               Keller, Alexander},
  title     = {{Sionna}: An Open-Source Library for
               Next-Generation Physical Layer Research},
  journal   = {arXiv preprint arXiv:2203.11854},
  year      = {2022}
}

@techreport{bib:3gpp38214,
  author      = {{3GPP}},
  title       = {{NR}; Physical Layer Procedures for Data},
  institution = {3rd Generation Partnership Project},
  type        = {Tech. Spec.},
  number      = {TS~38.214 V19.4.0},
  month       = jun,
  year        = {2026}
}

@techreport{bib:3gpp38321,
  author      = {{3GPP}},
  title       = {{NR}; Medium Access Control ({MAC}) Protocol Specification},
  institution = {3rd Generation Partnership Project},
  type        = {Tech. Spec.},
  number      = {TS~38.321 V19.3.0},
  month       = jun,
  year        = {2026}
}

@techreport{bib:3gpp38901,
  author      = {{3GPP}},
  title       = {Study on Channel Model for Frequencies
                 from 0.5 to 100~{GHz}},
  institution = {3rd Generation Partnership Project},
  number      = {TR~38.901 V17.0.0},
  month       = mar,
  year        = {2022}
}

@article{whitt1983qna,
  author  = {Whitt, Ward},
  title   = {The Queueing Network Analyzer},
  journal = {Bell System Technical Journal},
  volume  = {62},
  number  = {9},
  pages   = {2779--2815},
  year    = {1983},
  month   = nov,
  doi     = {10.1002/j.1538-7305.1983.tb03204.x}
}

@article{farzaneh2026intent,
  author  = {Amirmohammad Farzaneh and Salvatore D'Oro and Osvaldo Simeone},
  title   = {Should {I} Have Expressed a Different Intent? {Counterfactual} Generation for {LLM}-Based Autonomous Control},
  journal = {arXiv preprint arXiv:2601.20090},
  year    = {2026}
}

@article{hou2025counterfactual,
  author  = {Qiushuo Hou and Sangwoo Park and Matteo Zecchin and Yunlong Cai and Guanding Yu and Osvaldo Simeone},
  title   = {What If We Had Used a Different App? {Reliable} Counterfactual {KPI} Analysis in Wireless Systems},
  journal = {IEEE Transactions on Cognitive Communications and Networking},
  volume  = {11},
  number  = {5},
  pages   = {3529--3543},
  year    = {2025},
  doi     = {10.1109/TCCN.2025.3536793}
}

@article{prasad2026blinc,
  author  = {Reshma Prasad and Michele Polese and Tommaso Melodia},
  title   = {{BLINC}: Context-Specific Causal Learning for Automated {RAN} Configuration},
  journal = {arXiv preprint arXiv:2604.27084},
  year    = {2026}
}

@inproceedings{xue2026causaltune,
  author    = {Leyang Xue and Bolun Zhang and Yibo Ma and Mahesh K. Marina and
               He Yan and Yu Zhou and Cheuk Yiu Ip and Senthil Dhandapani and
               James Klosowski},
  title     = {{CausalTune}: Causal Learning Based Automated Cellular
               {RAN} Configuration Tuning Framework},
  booktitle = {Proc. of the {ACM SIGCOMM} 2026 Conference},
  address   = {Denver, CO, USA},
  month     = aug,
  pages     = {242--262},
  year      = {2026},
  doi       = {10.1145/3789240.3829109}
}

@article{vanderweele2009distinction,
  author  = {VanderWeele, Tyler J.},
  title   = {On the Distinction Between Interaction and Effect Modification},
  journal = {Epidemiology},
  volume  = {20},
  number  = {6},
  pages   = {863--871},
  year    = {2009},
  doi     = {10.1097/EDE.0b013e3181ba333c}
}

\end{document}